# Low-field anomalous magnetic phase in the kagomé-lattice shandite $Co_3Sn_2S_2$

Mohamed A. Kassem[*,†], Yoshikazu Tabata, Takeshi Waki, Hiroyuki Nakamura

Department of Materials Science and Engineering, Kyoto University, Kyoto 606-8501, Japan.

**Abstract**

The magnetization process of single crystals of the metallic kagomé ferromagnet $Co_3Sn_2S_2$ was carefully measured via magnetization and AC susceptibility. Field-dependent anomalous transitions observed in low fields indicate the presence of an unconventional magnetic phase just below the Curie temperature, $T_C$. The magnetic phase diagrams in low magnetic fields along different crystallographic directions were determined for the first time. The magnetic relaxation measurements at various frequencies covering five orders of magnitude from 0.01 to 1000 Hz indicate markedly slow spin dynamics only in the anomalous phase with characteristic relaxation times longer than 10 s.

1. **Introduction**

The kagomé lattice, a two-dimensional (2D) corner-sharing triangular-lattice, is a platform of many novel phenomena originating from its geometry. For instance, an antiferromagnetically coupled spin system on the kagomé lattice is highly frustrated and exhibits a quantum disordered spin-liquid state[1] and novel magnetically ordered states with nontrivial spin textures[2–5]. The geometric spin frustration is one of the most intriguing issues in recent condensed-matter physics[6–9]. The asymmetric Dzaloshinski–Moriya (DM) interaction is another driving force to nontrivial spin textures in the kagomé magnets[8–11]. The DM interaction can stabilize noncollinear ferro- and antiferromagnetic orders, resulting in spiral or canted spin structures. Particularly when the principal exchange interaction is ferromagnetic, a topologically-protected vortex-like spin texture, so-called 'skyrmion', may be realized by the DM interaction. Theoretically, spontaneous triangular-lattice formation of skyrimons was proposed in the presence of the DM interaction based on the phenomenological continuum model[12]. Indeed, the skyrmion lattices have been observed experimentally in noncentrosymmetric cubic chiral[13–17] and rotationally symmetric polar[18] crystals under finite fields and just below the Curie temperature, $T_C$. Even without the DM interaction, similar spin textures can be formed in uniaxial ferromagnets via anisotropic dipolar interaction[19]. Theoretically, a frustration-induced (non-DM-driven) skyrmion lattice in a centrosymmetric crystal has also been proposed[20]. Thus nontrivial spin textures in ferromagnetically coupled materials as well as novel magnetic state in the kagomé lattice now attract much attention.

$Co_3Sn_2S_2$ consists of Co-Sn metallic layers, stacked along $c$-axis in the hexagonal setting of $R\bar{3}m$ structure, that are separated by Sn-S blocks[21–26]. In each layer, Co atoms are arranged in a 2D kagomé sublattice as shown in Fig. 1. Previous studies of the magnetization process of the half metallic ferromagnet $Co_3Sn_2S_2$[26,27] have explored high axial anisotropy in temperature and field dependences of the magnetization $M(T, H)$[28,29]. The Curie temperature, $T_C$, and the spontaneous moment, $M_s$, at 2 K in the easy $c$-axis were reported as 174 K and ~0.3 $\mu_B$/Co, respectively. A recent comprehensive magnetization measurement and its analysis based on the spin fluctuation theory revealed the quasi-two-dimensionality (Q2D) of the magnetism in pure and In-substituted $Co_3Sn_2S_2$[30]. This implies that $Co_3Sn_2S_2$ is a model system of the 2D kagomé ferromagnet, and has a potential to exhibit novel magnetic states with nontrivial spin textures associated with the DM interaction and/or the strong frustration inherent to the kagomé lattice, if an antiferromagnetic component coexists, as well as the classical magnetic anisotropy in the uniaxial crystal.

In this paper, we present results of magnetization, $M(T, H)$, and AC susceptibility, $\chi_{ac}(T, H, f)$, of $Co_3Sn_2S_2$ systematically accumulated at low fields, and demonstrate the presence of a low-field anomalous but equilibrium magnetic phase just below $T_C$. The anomalous phase shows slow spin dynamics with characteristic relaxation times longer than 10 s, suggesting that the anomalous phase has a nontrivial spin texture.

**2. Experimental Details**

Single crystals of $Co_3Sn_2S_2$ used in this work were grown by a flux method[31,32]. Magnetization processes were measured using a SQUID magnetometer (MPMS, Quantum Design) under magnetic fields applied parallel and perpendicular to the easy *c*-axis. We confirmed reproducibility of magnetization anomalies mentioned below by measuring different crystals with various sizes and shapes. For AC susceptibility measurements, 1 Oe drive AC field of different frequencies, *f*, was superimposed on the bias field; both were applied along the *c*-direction. Two experimental regimes have been followed in the measurement of temperature scans: (i) Zero-field-cooled (ZFC) scans, after resetting field to zero, the sample was cooled to 5 K. Magnetic fields were applied at 5 K and the measurement was performed by increasing temperature stepwise and waiting for thermal equilibrium at each temperature. (ii) Field-cooled (FC) scans, in which the sample was cooled from 250 K, where the DC and/or AC magnetic fields were applied, and the measurement was performed with decreasing temperature stepwise.

To exclude the effect of domain wall motion, we measured low-field $\chi_{ac}$ of a single-domain crystal, namely we cooled the sample under the applied field of 1 kOe, higher than the saturation field, then reduced the field down to 50 Oe, and measured $\chi_{ac}$ in the heating process. We confirmed that, after this procedure, the DC magnetization keeps the saturated value and $\chi_{ac}$ reproduces the same characteristic anomalies observed in the DC magnetization at the same field, indicating that observed properties are intrinsic rather than those of non-equilibrium domain wall motion.

The spin relaxation processes at selected temperatures were systematically studied by measuring the ZFC AC susceptibility under different magnetic fields and in a frequency range of five orders of magnitude from 0.01 to 1000 Hz. In the frequency scans, the above ZFC regime was followed and fields were applied at the temperature of interest before the measurement.

**3. Results**

*3.1. Magnetization*

Typical temperature dependences of ZFC and FC magnetizations, $M(T, H)$, of a $Co_3Sn_2S_2$ single crystal measured at various low magnetic fields applied along and perpendicular to the *c*-axis are shown in Figs. 2(a) and (b), respectively. Both axial and in-plane magnetizations, $M_c$ and $M_{ab}$, increase rapidly at $T_C \approx$ 173 K, which corresponds well to reported ferromagnetic transitions[26,28,29,31]. As in literature, strong magnetic anisotropy was confirmed below $T_C$. When the magnetization is brought down after ZFC, the difference between ZFC and FC $M(T, H)$ is reduced with increasing magnetic field and disappears above ~1 kOe in the axial magnetization.

Interestingly, characteristic anomalies were found at low fields where the ZFC and FC magnetizations separate from each other. The axial magnetization, $M_c(T, H)$ in the FC process shows a hump at $T_A \approx 126$ K with applying $H = 5$ Oe, indicating the presence of another phase transition apart from the transition at $T_C$. The anomaly-temperature, $T_A$, slightly increases with increasing field. On the other hand, the ZFC magnetization shows a dip in the intermediate temperature range between $T_A$ and $T_C$. Both the hump in FC $M_c(T, H)$ and the dip in ZFC $M_c(T, H)$ disappear at $H \geq 400$ Oe. One can note that the anomalies are observed at fields below the saturation fields of $Co_3Sn_2S_2$[30].

The in-plane magnetization, $M_{ab}(T, H)$, displayed in Fig. 2(b), shows a similar hump at $T_A$ in the FC process and a dip between $T_A$ and $T_C$ in the ZFC process. Additionally, ZFC $M_{ab}(T, H)$ also shows a hump at $T_A$ at low fields ($\leq 150$ Oe) as seen in the inset of Fig. 2(b). Field-dependences of the anomalies in $M_{ab}$ are less pronounced than those in $M_c$. Especially, the dip in the ZFC process is still observable above 400 Oe.

### 3.2. AC susceptibility of $Co_3Sn_2S_2$

For more detailed investigation, we have carefully measured the temperature-, field- and frequency-dependences of the AC susceptibility, $\chi_{ac}(T, H, f)$, after ZFC and FC in a wide frequency range from 0.01 to 1000 Hz. Figure 3 shows the temperature dependence of the ZFC in-phase AC susceptibility, $\chi'(T)$, measured with $f = 1$ Hz and $H = 150$ Oe applied along the $c$-axis. For comparison, the ZFC magnetization, $M_c(T)$, measured at the same DC field is shown. $\chi'(T)$ shows distinct anomalies as in the DC magnetization; a sharp peak at $T_C$, a hump at $T_A$, and a dip between $T_A$ and $T_C$.

Figure 4 shows typical examples of temperature dependences of ZFC and FC real and imaginary parts of the AC susceptibility, $\chi'(T, H)$ and $\chi''(T, H)$, at under 1 Oe AC fields of $f = 1, 10, 100,$ and 1000 Hz, and DC fields of 0, 150, and 600 Oe applied along the $c$-axis. The characteristic features in $\chi'$ are seen in Fig. 3 ($f = 1$ Hz and $H = 150$ Oe) and Figs. 4(a) and (c) for all the frequencies and the fields of 0 and 150 Oe. At 600 Oe (Fig. 4(e)), only the anomaly at $T_C$ is seen. The hump and the broad dip are absent in the DC magnetization at the fields higher than 400 Oe (Fig. 2). At 600 Oe, the anomaly at $T_C$ is rather broad, suggesting different natures at the low and high fields. It should be noted that $\chi'$ in ZFC and FC processes almost coincide with each other at the low and high fields, unlike the DC magnetization.

The characteristics at low fields are more pronounced in the imaginary part of the AC susceptibility, $\chi''$, as seen in Figs. 4(b) and (d). $\chi''$ is almost absent above $T_C$ and shows a rapid increase at $T_C$. The anomalies below $T_C$ are not so distinct as in $\chi'$ except for that at 1 Hz. It should be more emphasized that $\chi''$ is enhanced appreciably only in the particular temperature range between $T_A$ and $T_C$ and in the low-field range less than 150 Oe. At 600 Oe, $\chi''$ is almost absent regardless of temperature and frequency, as shown in Fig. 4(f). The appreciable magnitude of $\chi''$ in the frequency region of 1–1000 Hz indicates that spin fluctuations slow down to the time range of 0.001–1 s only between $T_A$ and $T_C$ and at the low fields. This gives a clear evidence of the emergence of the anomalous phase apart from a conventional ferromagnetic state.

### 3.3. Magnetic phase diagrams

Here we present magnetic phase diagrams of $Co_3Sn_2S_2$ based on the magnetization and AC susceptibility data described above. Figure 5(a) illustrates the $H$–$T$ phase diagram when magnetic field is applied along the $c$-axis, where $T_C$ and $T_A$ were determined from the FC magnetization and the ZFC AC susceptibility. $T_A$ is assigned from the hump in the FC magnetization and ZFC $\chi'$, and $T_C$ from the sharp peak in the FC magnetization, ZFC $\chi'$, and ZFC $\chi''$. The $H$-$T$ region surrounded by $T_A$ and $T_C$ is denoted as the A phase. In Fig. 5(a), the saturation field, $H_s$, where the magnetization saturates to the high-field limit, is also plotted. It should be noted that $H_s(T)$ merges the high-field boundary of $T_A(H)$ at the A phase, suggesting that the A phase is characterized by a nontrivial spin texture prior to a field-stabilized collinear ferromagnetic structure.

Figure 5(b) illustrates the $H$–$T$ phase diagram when magnetic field is applied perpendicular to the $c$-axis. The A phase boundary is not closed because the in-plane magnetization does not saturate due to the high magnetic anisotropy.

### 3.4. Relaxation phenomena and frequency dependences of $\chi'$ and $\chi''$.

The existence of non-zero $\chi''$ in the A phase, particularly between $T_C$ and $T_A$, as described in Sec. 3.2, implies that spin fluctuations slow down to the experimental time window in the temperature range. To see the slow spin dynamics in the A phase in more details, frequency dependences of the AC susceptibility were measured at temperatures which are outside (lower) and inside of the A phase.

ZFC $\chi'$ and $\chi''$ at several temperatures and under $H = 150$ Oe are displayed in Figs. 6(a) and (b), respectively. $\chi'$ below and at around $T_A$ are weakly frequency-dependent, and correspondingly, $\chi''$ are almost absent. The increasing trend of $\chi''$ at high frequencies can be ascribed to an appearance of frequency dependence in the time window, which has been found only in the MHz range in ferromagnetic materials[33], or to an extrinsic effect due to eddy currents at low temperatures. While the boundary from the ferromagnetic (FM) phase to the A phase, $T_A$, was evaluated from the hump in $\chi'$, as described in Sec. 3.2, it is not appreciable in Figs. 6(a) and (b). $\chi'$ and $\chi''$ are featureless up to 135 K, where a minimum is seen. At 160 K, deeper inside the A phase, $\chi'$ and $\chi''$ exhibit pronounced frequency-dependences. $\chi''$ increases strongly down to 0.5 Hz, indicating a characteristic time scale of spin dynamics longer than 2 s at 160 K.

Frequency dependences of $\chi'$ and $\chi''$ becomes more pronounced with approaching the high-temperature boundary between the A and paramagnetic phases as shown in Figs. 6(c) and (d). It is clearly seen that the time scale of spin dynamics is reduced with increasing temperature and exceeds the experimental time window (> 173 K) above $T_C$. Frequency dependences of $\chi'$ and $\chi''$ just below zero-field $T_C$, 172 K, are shown in Figs. 6(e) and (f) for different magnetic fields. Apart from the magnitude, the overall shape of frequency dependences is not modified significantly at $T \sim T_C$, indicating weak field variation of the spin dynamics at $T_C$. The magnitudes of both $\chi'$ and $\chi''$ are suppressed with approaching the high-field boundary of the A phase, ~200 Oe, above which both $\chi'$ and $\chi''$ are almost frequency-independent. These results can be interpreted as that, the field fixes spin along its direction, and the higher the field, spins fluctuate more slowly.

To see the evolution of spin dynamics in the A phase, the phenomenological Cole-Cole formalism taking account of the distribution of spin relaxation times[34,35] is employed. In the formalism, the AC susceptibility is given by

$$\chi(\omega) = \chi(\infty) + \frac{A_0}{1 + (i\omega\tau_0)^{1-\alpha}}, \qquad (1)$$

with $A_0 = \chi(0) - \chi(\infty)$, where $\chi(0)$ and $\chi(\infty)$ are the isothermal and adiabatic susceptibilities, $\omega = 2\pi f$ is the angular frequency, $\tau_0$ is the relaxation time, and $\alpha$ is the parameter which provides a measure of the distribution of the relaxation time. $\alpha = 0$ reverts Eq. (1) to the conventional Debye relaxation with a single relaxation time, and $\alpha = 1$ gives an infinite width of the distribution. Real and imaginary parts of the AC susceptibility can be extracted from Eq. (1) as[34–36]

$$\chi'(\omega) = \chi(\infty) + \frac{A_0 \left[1 + (\omega\tau_0)^{1-\alpha} \sin(\pi\alpha/2)\right]}{1 + 2(\omega\tau_0)^{1-\alpha} \sin\left(\frac{\pi\alpha}{2}\right) + (\omega\tau_0)^{2(1-\alpha)}}, \qquad (2)$$

$$\chi''(\omega) = \frac{A_0 (\omega\tau_0)^{1-\alpha} \cos(\pi\alpha/2)}{1 + 2(\omega\tau_0)^{1-\alpha} \sin\left(\frac{\pi\alpha}{2}\right) + (\omega\tau_0)^{2(1-\alpha)}}. \qquad (3)$$

The fitting results of experimental data in the vicinity of $T_C$ to Eqs. (2) and (3) are shown by solid curves in Figs. 6(c), (e) and (d), (f), respectively. The fitting parameters, $\alpha$ and $f_0$, are shown as functions of temperature at $H = 150$ Oe and as functions of the applied field at $T = 172$ K in the insets of Figs. 6(d) and (f), respectively. Estimated $\alpha$, little varying at around 0.6 against $T$ and $H$, indicates a significant distribution of relaxation times and its weak temperature and field variations. On the other hand, the characteristic frequency, $f_0 = 1/\tau_0$, varies. Especially, $f_0$ drastically decreases with decreasing temperature at around $T_C$, i.e., $f_0 \approx 25$ Hz at 172 K, $f_0 \approx 0.1$ Hz at 171 K, and $f_0$ is lower for $T = 170$ K. The field variation of $f_0$ at $T_C$ is more moderate, as seen in the inset of Fig. 6(f).

It should be noted that, at the $T$- and $H$-regions corresponding to the A phase, bell-shaped frequency dependences of $\chi''$ are asymmetric as seen in Figs. 6(d) and (f), implying that the Cole-Cole formalism in not valid at low frequencies. This suggests a more complicated mechanism of spin fluctuations in the A phase.

**4. Discussion**

The magnetization and the AC susceptibility suggest zero- and low-field characteristic magnetic transitions to a nontrivial spin state just below $T_C$. The magnetic relaxation with a characteristic time scale of several seconds at around the phase boundaries and the slower relaxation inside the A phase indicate that the A phase is a kind of spin-frozen state. The slow spin dynamics can primarily be ascribed to spin-glass-like state[37]. However, no disorder in our system and, moreover, with the fast relaxation of a few microseconds in the ferromagnetic state below $T_A$ excludes the possibility of the glass-like state. Non-equilibrium phenomena such as domain-wall motions have been excluded to be the origin of these anomalous magnetic transitions, as described in section 2.

The presence of the DM interaction and the strong geometric frustration is inherent to the kagomé lattice. Hence, the DM interaction and/or spin frustration, if antiferromagnetic component of the exchange interaction coexists, may stabilize noncolinear spin structures. Multi-**q** states like magnetic skyrmions may be realized in the rotationally symmetric lattice. Classically, an anisotropic dipolar interaction may result in similar skyrmion-like states in the uniaxial lattice. If topologically protected skyrmions were realized, we expect slow spin dynamics because of their high stability. In chiral skyrmion hosts[36,38–40], slow spin dynamics in the time scale of $\chi_{ac}$ experiments and the distribution of relaxation times near $T_C$ have actually been observed.

In contrast to chiral and polar skyrmion hosts[18,41], where the A phases are observed only at finite fields, the A phase in $Co_3Sn_2S_2$ is extended to zero field. In centrosymmetric lattices with additional degrees of freedom, a richer variation of spin textures, including stripe domains, soft and hard magnetic bubbles, biskyrmions, etc., can be realized[42]. One of them may explain the nature of the A phase in $Co_3Sn_2S_2$. Further small-angle neutron scattering (SANS) experiments, spin-polarized scan tunneling microscopy (STM) or Lorentz TEM microscopy are helpful to shed light on the nature of the magnetic states of $Co_3Sn_2S_2$.

5. **Conclusion**

Precise measurements of the magnetization and the AC susceptibility for the kagomé ferromagnet $Co_3Sn_2S_2$ provided an approach to its low-field *H-T* phase diagrams. The presence of a zero- and low-field anomalous equilibrium phase just below $T_C$, was established for the first time. The frequency-distributed spin relaxation process with characteristic relaxation times of the order of several seconds has been observed at the A phase boundary. The microscopic nature of the novel magnetic state should be revealed by further microscopic experiments.

**Acknowledgement**

The authors would like to thank I. Kezsmarki from Budapest University of Technology and Economics for valuable discussion on the skyrmion. M.A.K. thanks the Egyptian Ministry of Higher Education for financial support during his study in Kyoto University, Japan. This work was partly supported by Murata scientific foundation.

**Figures captions**

**Figure 1:** Crystal structure projection in (001) plane, drawn using VESTA software[43], shows a Co kagomé sublattice in $Co_3Sn_2S_2$.

**Figure 2**: Typical temperature dependences of ZFC (open) and FC (closed) magnetizations, $M(T, H)$, of $Co_3Sn_2S_2$ measured at the indicated low magnetic fields applied (a) along and (b) perpendicular to the *c*-axis. Vertical arrows indicate the transition temperature, $T_A$, and the broad dip (see text). The inset of (b) shows a magnification of the ZFC $M(T)$ measured at very low fields.

**Figure 3:** Temperature dependences of ZFC $\chi'$ (at $f = 1$ Hz, on left axis) and $M$ (on right axis) of $Co_3Sn_2S_2$ at 150 Oe applied along the *c*-axis. Vertical dotted lines indicate $T_A$ (see text) and $T_C$.

**Figure 4:** Thermal variation of ZFC and FC $\chi'$ and $\chi''$ of $Co_3Sn_2S_2$ measured at different frequencies from 1 to 1000 Hz and for magnetic fields of zero (a), (b), 150 Oe (c), (d), and 600 Oe (e), (f), applied along the *c*-axis. Arrows indicate $T_C$ and $T_A$ (see text). For clarity purpose, the data with frequencies of 1, 10, and 100 Hz have been shifted vertically with respect to the baseline by multiplying their data to the numbers indicated as in (a) and (b).

**Figure 5:** Magnetic *H-T* phase diagrams of $Co_3Sn_2S_2$ for fields applied (a) along and (b) perpendicular to the *c*-axis. Dashed and solid curves are for the eye guidance.

**Figure 6**: Frequency dependences of ZFC-$\chi'$ and $\chi''$ of $Co_3Sn_2S_2$ at $T = 50–160$ K (a), (b) and at 170–172 K (close to $T_C$) (c), (d), under a magnetic field of 150 Oe applied along the *c*-axis. (e) and (f) show ZFC-$\chi'$ and $\chi''$ at $T_C \approx 172$ K for different fields up to 250 Oe applied along the *c*-axis. Solid lines in (c)–(f) represent fittings to Eqs. (2) and (3). Insets of (d) and (f) show the fitting parameters, $\alpha$ and $f_0$, as functions of temperature at $H = 150$ Oe and as functions of applied field at $T = 172$ K, respectively.

**Figures**

**Figure 1:**

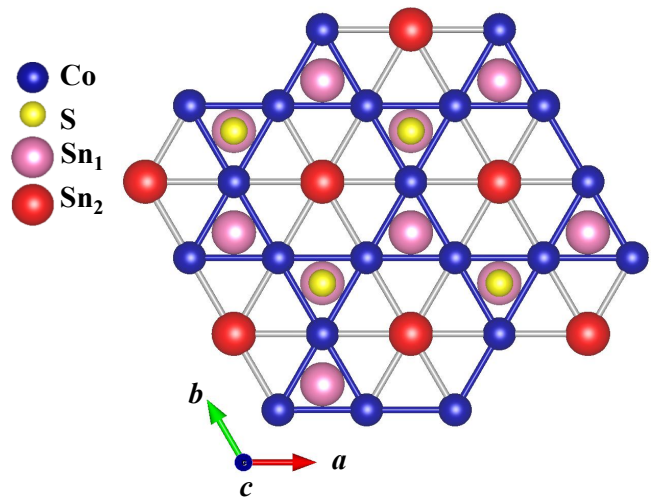

Figure 2:

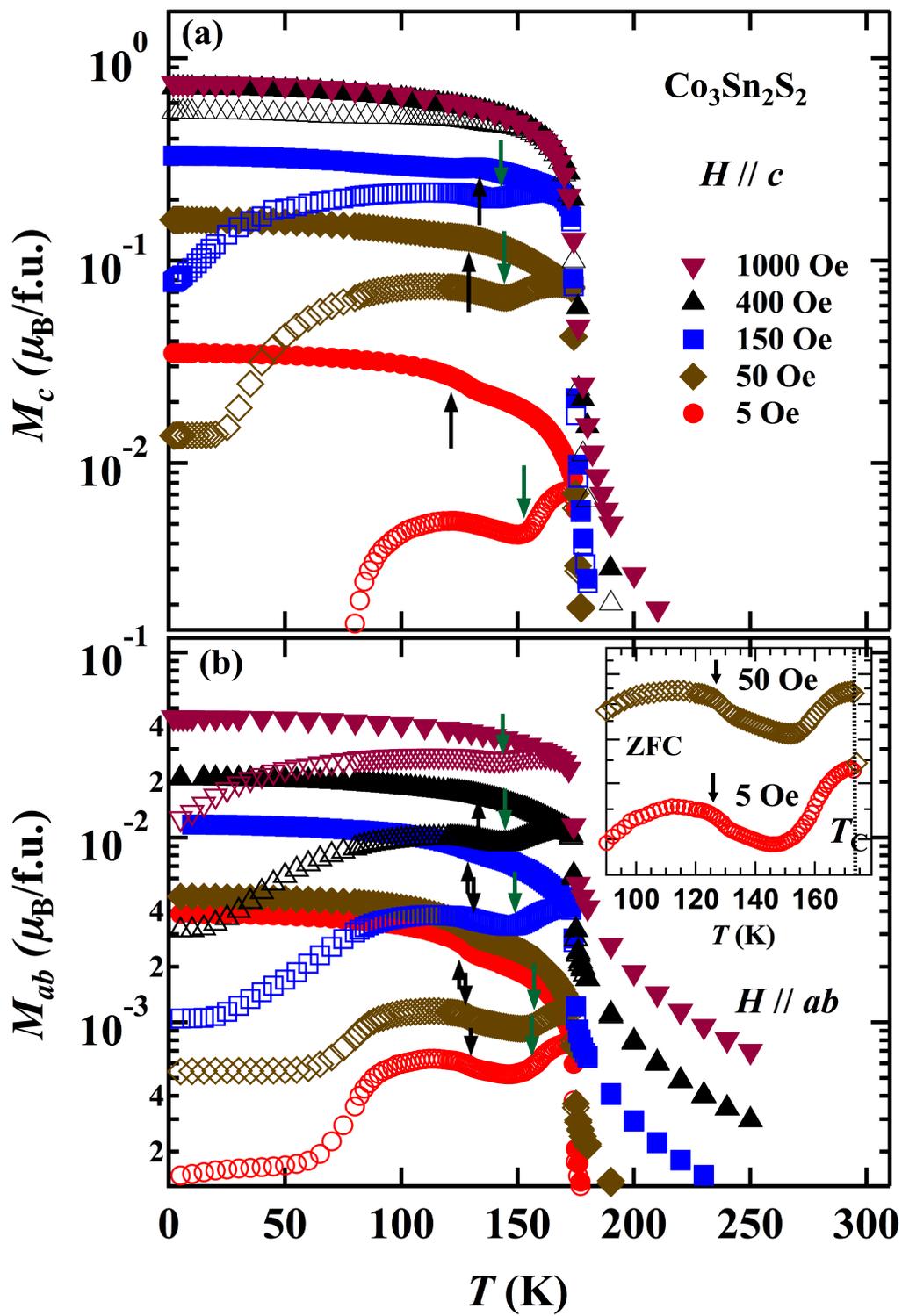

Figure 3:

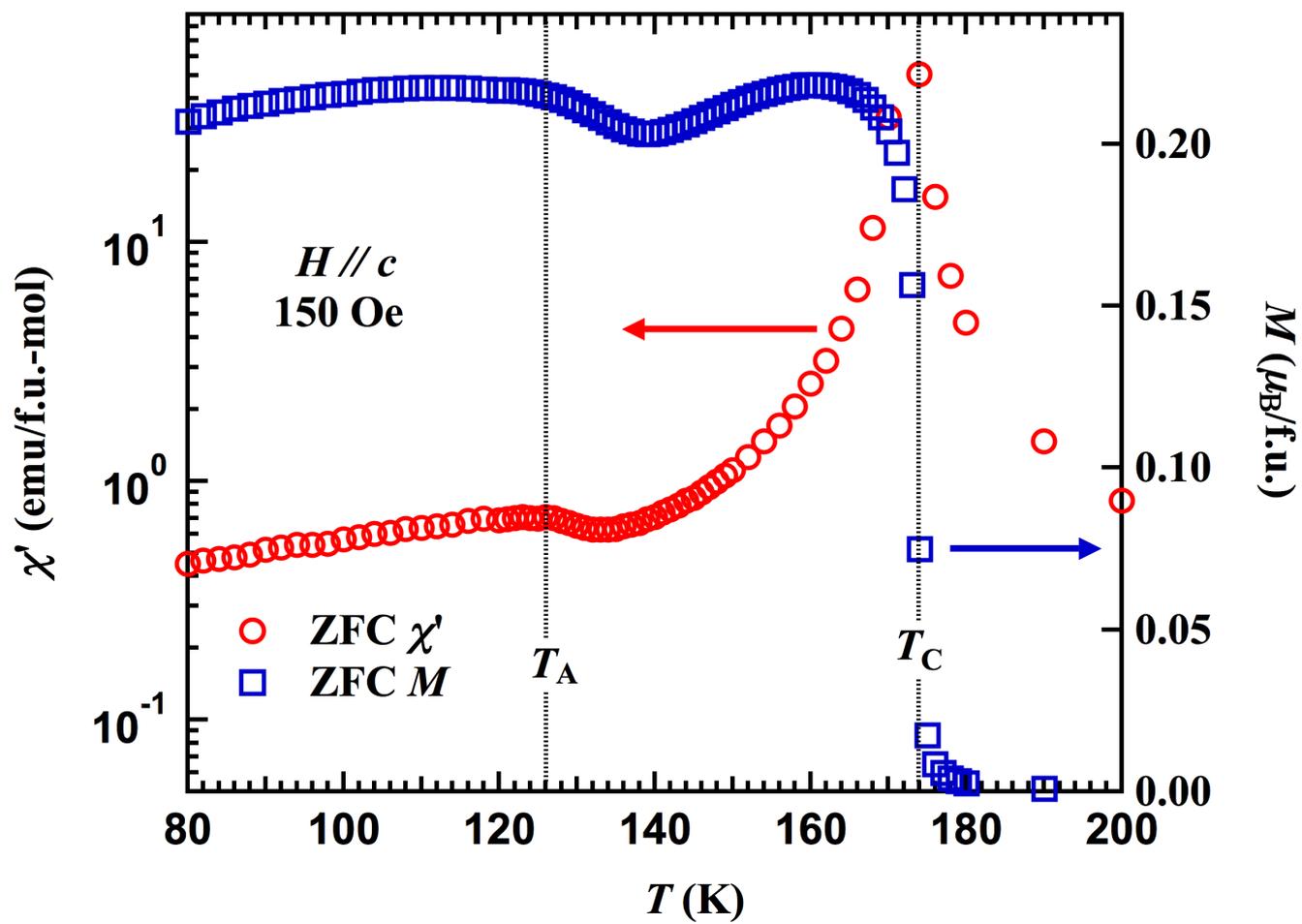

Figure 4:

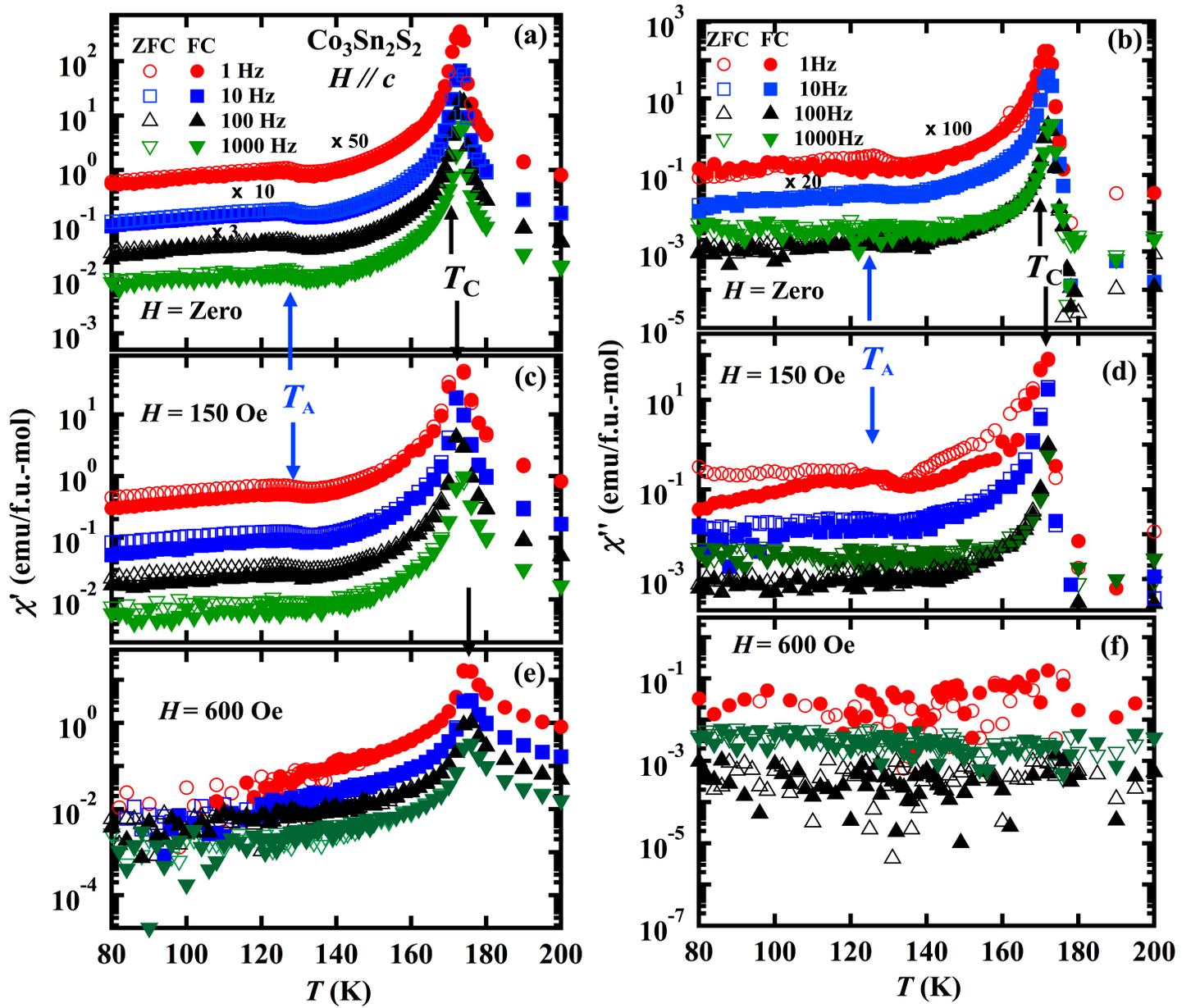

**Figure 5:**

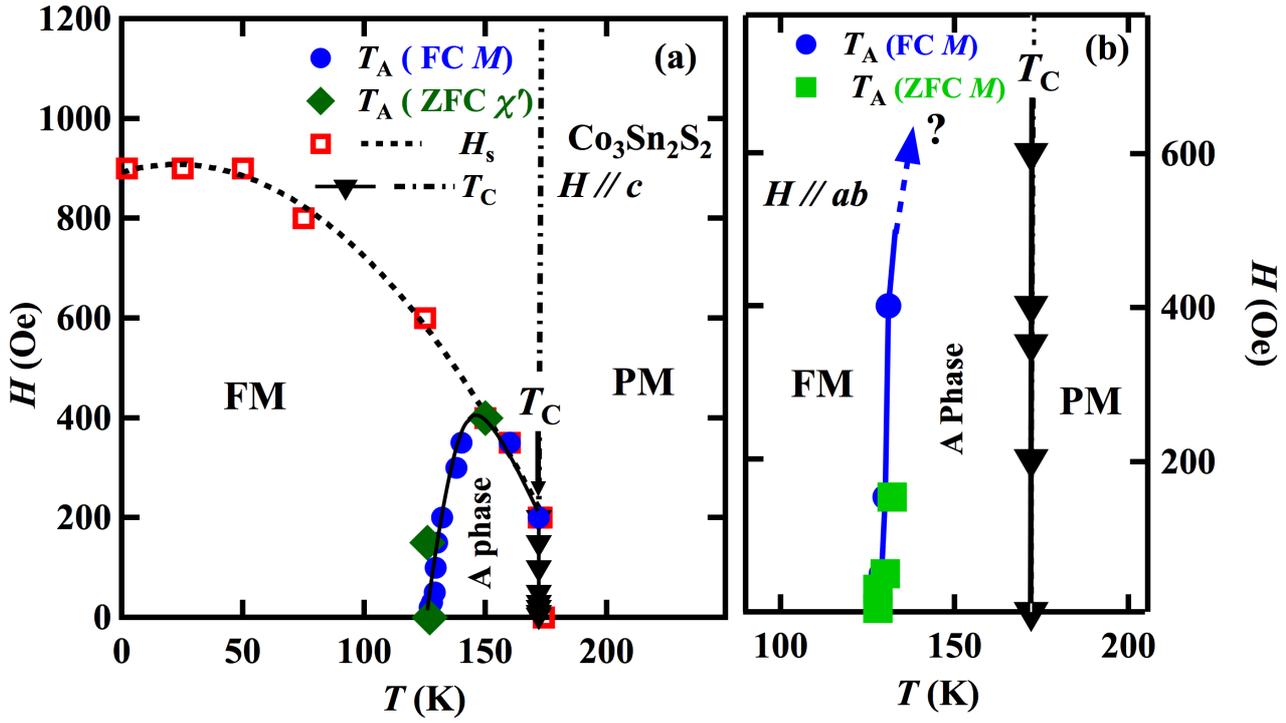

**Figure 6:**

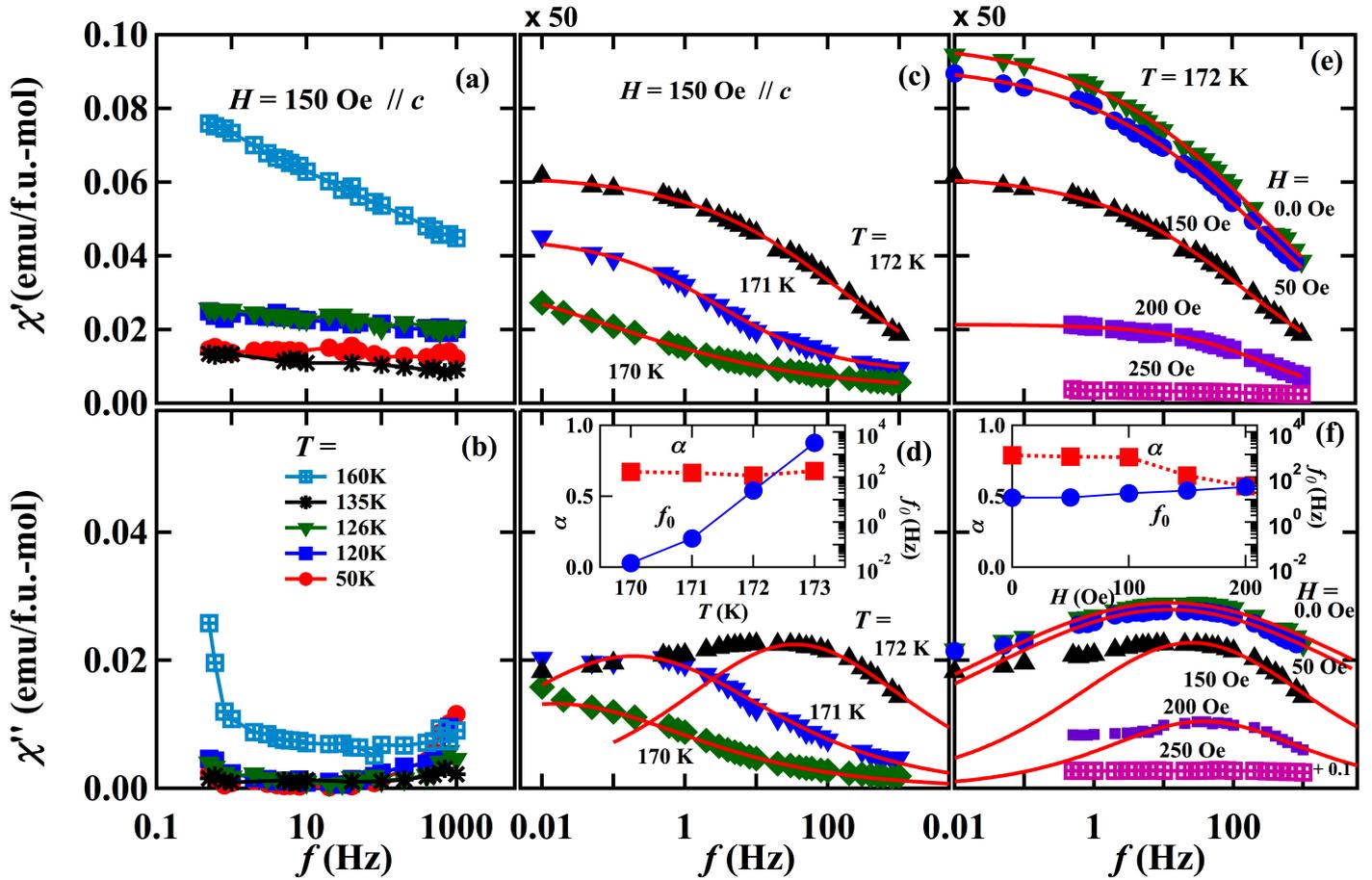

**Correspondence to:**
* **E-mail:** makassem@aun.edu.eg,
† **Permanent Address:** Department of Physics, Assiut University, Assiut 71516, Egypt.